\documentclass{article}
\usepackage{graphicx}
\usepackage[english]{babel}

\title{
\includegraphics[width=0.35\linewidth]{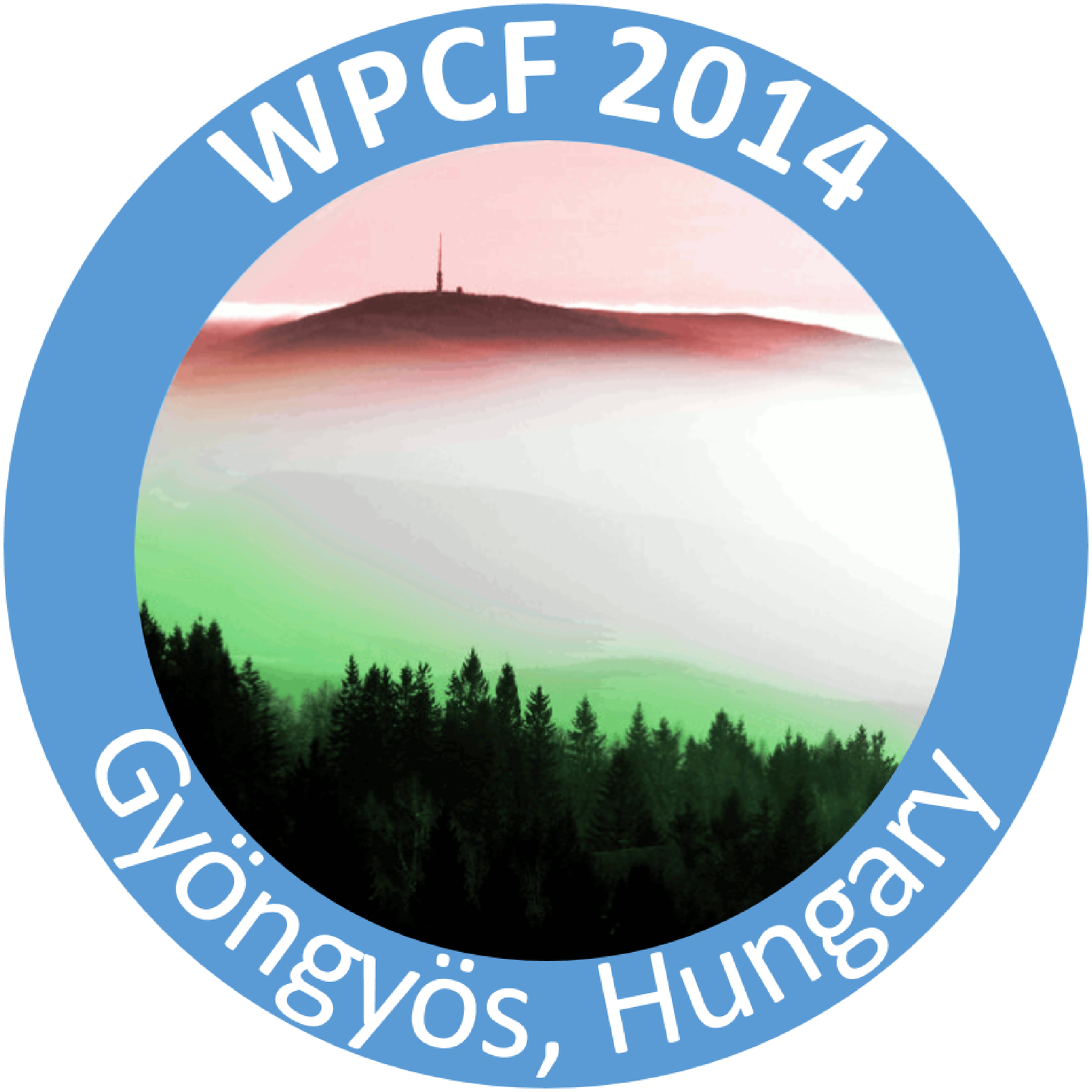}\\[1cm]
Study of in-medium mass modification at J-PARC}
\author{
K.~Aoki$^1$ for the J-PARC E16 Collaboration\\[1ex]
$^1$KEK, High Energy Accelerator Research Organization,\\
Tsukuba, Ibaraki 305-0801, Japan\\
}

\begin{document}
\fontfamily{lmss}\selectfont
\maketitle

\begin{abstract}
Study of in-medium mass modification has attracted 
interest in terms of the restoration of the spontaneously broken chiral symmetry,
which is responsible for the generation of hadron mass.
Many experiments were performed to measure in-medium property of hadrons
but there is no consensus yet.
J-PARC E16 has been proposed to study in-medium property of vector mesons
via dilepton decay channel.
The status of spectrometer R \& D is explained.
Other related experiments planned at J-PARC are also introduced.
\end{abstract}

\section{Introduction}
Spontaneous breaking of the chiral symmetry is considered
to be the origin of hadron mass.
The chiral symmetry is expected to be (partially) restored
in finite density and the hadron mass is predicted to decrease,
even at the normal nuclear density.
Our purpose is to investigate the origin of hadoron mass
through mass modification of hadrons.

\begin{figure}[hb]
\begin{center}
\includegraphics[width=.8\linewidth]{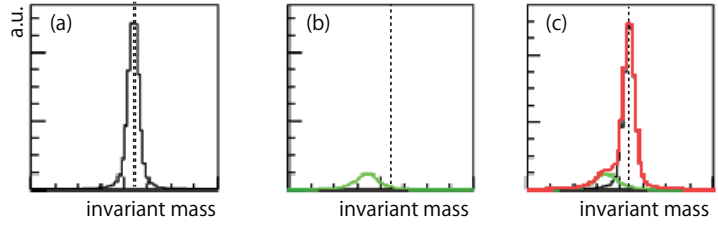}
\end{center}
\caption{Invariant mass spectra of $\phi$ meson (a) in vacuum , and (b) in medium.
(c) is the sum of (a)+(b). The dashed line indicates the $\phi$ meson mass
in vacuum.}
\label{fig:inv}
\end{figure}

The dilepton decay channel of vector mesons produced in nuclear reactions
is a good probe of in-medium mass modification
since it is free from the final state interactions.
We take $p+A \rightarrow \phi + X$ reaction as an example
to explain the expected invariant mass distribution of the
vector meson. A $\phi$ meson produced inside a target nucleus
travels and then decays inside or outside the target nucleus.
When the $\phi$ meson decays outside the target nucleus,
the mass spectrum is well-known invariant mass in vacuum
as in Fig.~\ref{fig:inv}(a).
When the $\phi$ meson decays inside the target nucleus,
the observed mass is the one in medium.
So if in-medium mass modification exists,
the mass distribution is modified to some extent as in Fig.~\ref{fig:inv}(b).
What we experimentally measure is the sum of these cases as in Fig.~\ref{fig:inv}(c).

\begin{figure}
\begin{center}
\begin{minipage}{0.40\linewidth}
\vspace{2ex}
\includegraphics[width=.95\linewidth]{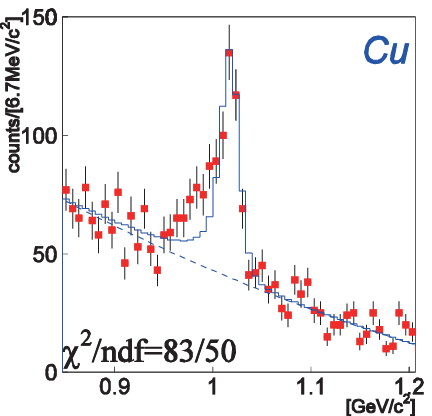}
\vspace{1ex}
\caption{Invariant mass spectrum of $\phi$ meson obtained with $p+$Cu reactions
by KEK E325 experiment. $\beta\gamma$ of $\phi$ is $<1.25$.
}
\label{fig:invphie325}
\end{minipage}
\begin{minipage}{0.10\linewidth}
~
\end{minipage}
\begin{minipage}{0.40\linewidth}
\includegraphics[width=.95\linewidth]{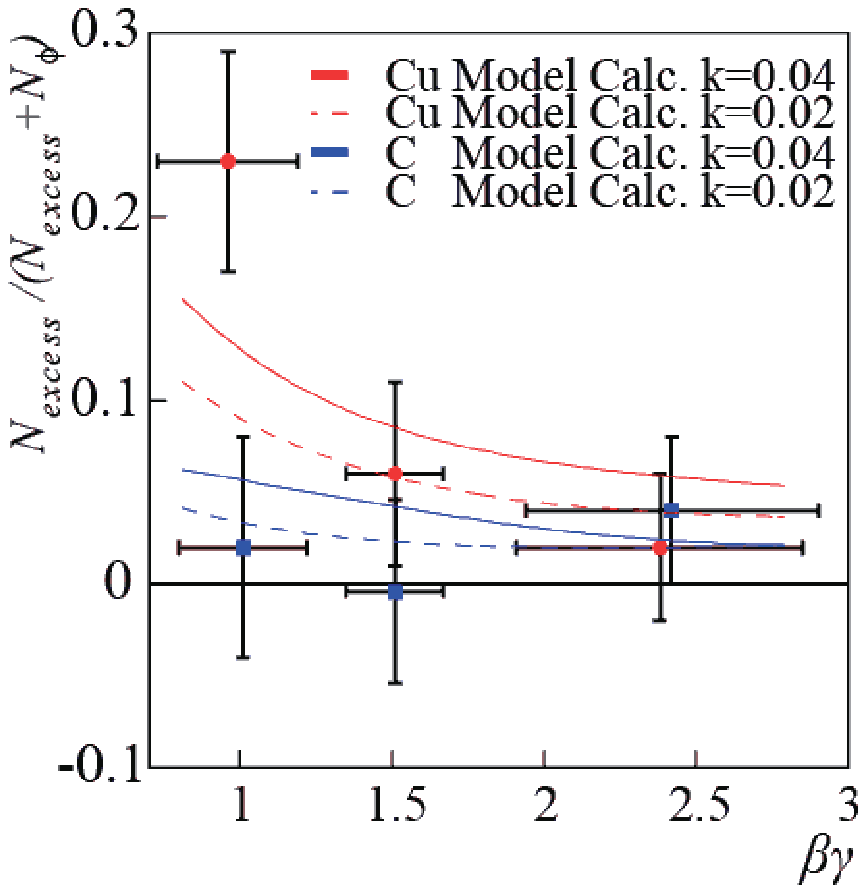}
\caption{Amount of excess versus $\beta\gamma$ of $\phi$
measured by KEK E325 experiment.}
\label{fig:bg-excess-e325}
\end{minipage}
\end{center}
\end{figure}

The KEK E325 experiment was performed at KEK Proton Synchrotron
to search for in-medium mass modification using the method explained above.
They measured the invariant mass spectra of $e^+e^-$ pairs produced
in 12~GeV proton beam induced nuclear reactions.
As the nuclear targets, C and Cu were used.
The mass resolution was about 11~MeV$/c^2$.
Figure~\ref{fig:invphie325} shows the invariant mass spectrum of $\phi$ meson
produced in Cu target with $\beta\gamma(=P/M)<1.25$~\cite{muto}. The blue line
represents an expected line shape assuming mass in vacuum
including experimental effects.
There is an excess on the lower side of the $\phi$ mass peak over the expected line shape.
Figure~\ref{fig:bg-excess-e325} shows the amount of excess versus $\beta\gamma$ of
$\phi$ mesons.
This figure support the picture that slower $\phi$ meson
in larger nuclear target have higher
probability to decay inside the nuclear target so that it experience medium modification.
To quantitatively extract information on the medium effect,
they assume linear dependence of the mass on density as,
\begin{equation}
\frac{m(\rho)}{m(0)} = 1 - k \frac{\rho}{\rho_0},
\end{equation}
where $m(\rho)$ is the mass at density $\rho$, $\rho_0$ is the normal nuclear density,
and $k$ is the parameter to be determined.
Similarly for the width, they assume
\begin {equation}
\frac{\Gamma(\rho)}{\Gamma(0)} = 1 + k_2 \frac{\rho}{\rho_0},
\end{equation}
where $\Gamma(\rho)$ is the width at density $\rho$, and $k_2$
is the parameter to be determined.
They obtained $k=0.034^{+0.006}_{-0.007}$ and $k_2=2.6^{+1.8}_{-1.2}$,
which means that the mass of $\phi$ decreases by 3.4\% and the width
gets wider by 3.6 times at the normal nuclear density.
They also observed modification of $\rho$ and $\omega$ mass and concluded that
$k=0.092 \pm 0.002$, assuming that the parameter is common for $\rho$ and
$\omega$~\cite{e325-omega}.
Width broadening was not necessary to reproduce the observed invariant mass.
The mass shift parameters $k$ for $\rho/\omega$, and $\phi$ are
at the same level as the calculations based on QCD sum rule~\cite{QCDsum}

The CLAS g7 experiment at Jefferson Laboratory
used $\gamma + A$ reactions and the light vector mesons
were reconstructed using $e^+e^-$ decay channel~\cite{g7}.
The results obtained with $^2$H, C, Fe, Ti targets were presented.
For $\omega$ and $\phi$ meson, no mass shift was assumed in the analysis
due to their long life. The $\omega$ and $\phi$ contributions were subtracted
to extract the invariant mass spectra of $\rho$ meson.
The mass of $\rho$ meson in nuclear medium does not show any mass shift.
The width are broadened and
it is consistent with an expectation from collisional broadening.

Dilepton invariant mass spectra were also measured in heavy ion collisions.
CERES /NA45 reported $e^+e^-$ invariant mass measured in 158~AGeV Pb+Au collisions~\cite{na45}.
An in-medium broadening of the $\rho$ mass scenario is
favored over a $\rho$ mass dropping scenario.
PHENIX also reported invariant mass spectra of $e^+e^-$ in Au+Au collisions
at $\sqrt{s}=200$A~GeV~\cite{phenix}.
An enhancement is observed in the low mass region (below $\phi$ peak).
The enhancement at quite low mass ($m_{ee}<0.3$~GeV$/c^2$) and
high $p_T$ ($1<p_T<5$~GeV/$c$)
is interpreted as the production of virtual direct photons,
which leads to their temperature measurement.
No theoretical models could explain quantitatively the enhancement
at the low mass and low $p_T$ region.

\section{J-PARC E16 Experiment}
There exists some modification in $e^+e^-$ mass spectrum but
the origin is not yet clear. There are even contradiction in the interpretation.
We propose to pursue this problem using the same
reaction as KEK-E325 but with 100 times more statistics ($10^3~\phi \rightarrow 10^5~\phi$)
and with two times better mass resolution (11~MeV/$c^2$ $\rightarrow$ 5~MeV/$c^2$). 
The proposal was approved as stage-1 and the experiment
was named E16~\cite{proposal}.
We use 30~GeV $p+A \rightarrow \rho/\omega/\phi\ X$ reactions 
and measure dilepton invariant mass spectra. As the nuclear target,
CH$_2$, C, Cu, Pb are used.
The J-PARC E16 experiment has the following advantages and disadvantages compared
to other experiments. It observes $e^+e^-$ decay channel so
it can eliminate final state interactions in contrast to the case of
experiments using hadronic decay channels.
However, $e^+e^-$ decay channel has very tiny branching ratio ($\sim 3\times 10^{-4}$
for $\phi$ meson).
The E16 experiment uses proton induced reactions,
therefore, the system is cold and static so is simpler compared to that of heavy ion collisions.
The E16 experiment is expected to measure $\phi$ meson invariant mass modification.
Compared to $\rho$ and $\omega$ mesons, $\phi$ meson has a non-overlapping
separated peak and a narrower width.
However, the production cross section of $\phi$ meson is much smaller 
than that of $\rho$ and $\omega$,
thus it is difficult to collect high statistics and
CLAS-g7 and CERES cannot discuss the mass spectra of $\phi$.
The disadvantages, mainly come from the fact
that $\phi \rightarrow e^+e^-$ is a rare probe,
are overcome by collecting high statistics data.

\begin{figure}
\begin{center}
\begin{minipage}{0.45\linewidth}
\includegraphics[width=.9\linewidth]{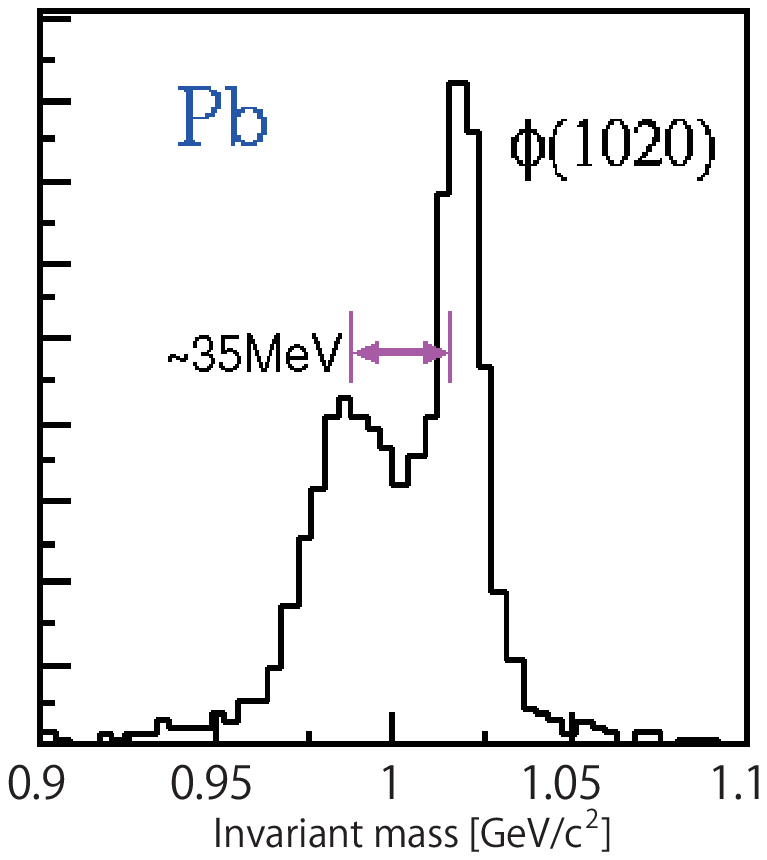}
\caption{Invariant mass distribution of $\phi$ meson with $\beta\gamma<0.5$
expected to be observed by J-PARC E16 experiment using Pb target.}
\label{fig:invmass-phi-e16}
\end{minipage}
\begin{minipage}{0.05\linewidth}
~
\end{minipage}
\begin{minipage}{0.45\linewidth}
\includegraphics[width=.9\linewidth]{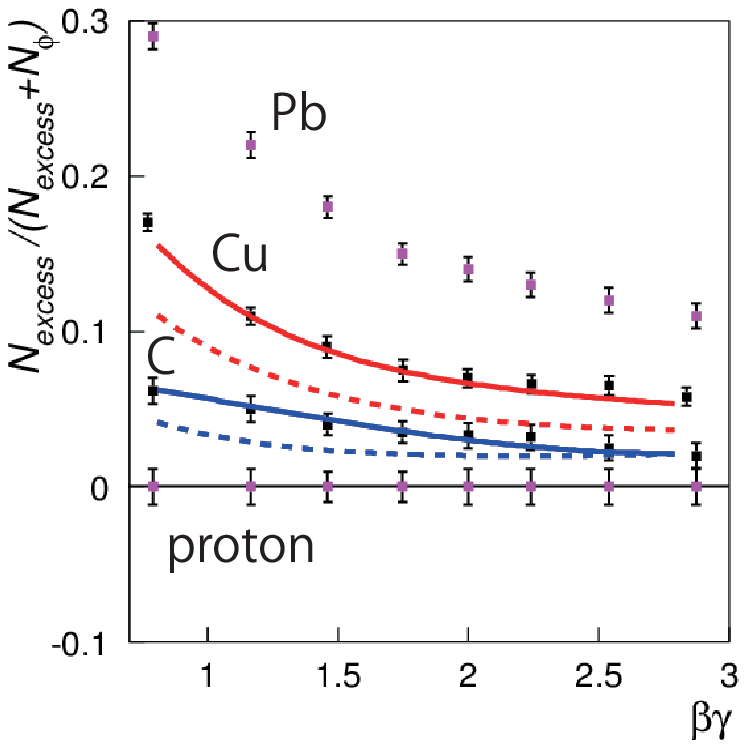}
\caption{Amount of excess versus $\beta\gamma$ of $\phi$
expected to be obtained by J-PARC E16 experiment.}
\label{fig:bg-excess-e16}
\end{minipage}
\end{center}
\end{figure}

\begin{figure}
\begin{center}
\includegraphics[width=.35\linewidth]{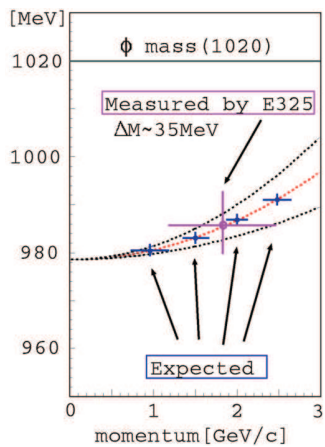}
\caption{Dispersion relation.
The red dotted curve shows a theory calculation by S.H.~Lee~\cite{lee}.
Note that the calculation is limited for momentum of less than 1~GeV/$c$
and is extrapolated to 3~GeV/$c$.
Black dotted curve shows the uncertainty of the calculation.
Blue points shows the statistical uncertainties
expected to be obtained by the J-PARC E16 experiment.
The center values are taken 
from the theoretical calculation mentioned above.
Purple point is the results obtained by KEK E325 experiment.
}
\label{fig:dispersion}
\end{center}
\end{figure}

When the statistics is achieved,
the invariant mass distribution of slowly moving $\phi$ meson whose $\beta\gamma$ 
is less than 0.5 which is obtained with Pb target is expected to have double peak
as in Fig.~\ref{fig:invmass-phi-e16}.
Note that the modification parameters obtained by KEK-E325 are assumed.
The $\beta\gamma$ and the target size dependence of the modification
expected to be obtained is in Fig.~\ref{fig:bg-excess-e16}.
So more systematic study is possible.
We are able to obtain dispersion relation as the blue points in Fig.~\ref{fig:dispersion},
which is qualitatively new information.
These new information can give further insight on the in-medium modification.

\subsection{J-PARC and the high momentum beam line}
To achieve 100 times more statistics, We utilize 10 times more intense beam
($10^9$ protons per pulse (ppp) $\rightarrow$ $10^{10}$ ppp),
a spectrometer with 5 times larger acceptance,
and 2 times larger production rate due to the increased beam energy
(12~GeV $\rightarrow$ 30~GeV).

\begin{figure}
\begin{center}
\includegraphics[width=0.95\linewidth]{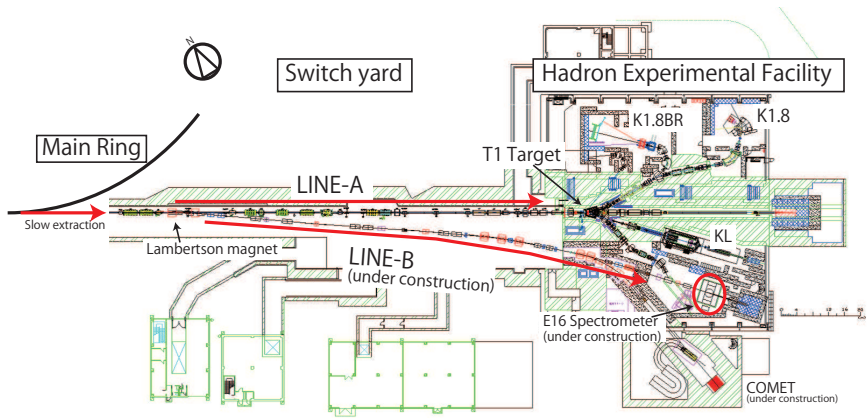}
\caption{Plan view of the switchyard and the Hadron Experimental Facility.}
\label{fig:sy-hd}
\end{center}
\end{figure}

The J-PARC E16 experiment plans to use the high momentum beam line
which will be constructed at J-PARC Hadron Experimental Facility.
J-PARC, Japan Proton Accelerator Research Complex,
is a high intensity proton accelerator and is located at Tokai village in Japan.
The Main Ring (MR) of J-PARC can accelerate protons up to 30~GeV.
Figure~\ref{fig:sy-hd} shows the plan view of the switchyard and
the Hadron Experimental Facility.
The protons in the MR are slowly extracted to LINE-A.
The proton beam follows LINE-A through the switchyard
and is delivered to the Hadron Experimental Facility.
The protons collide the T1 target to provide secondary beams
to the existing beam lines such as K1.8, K1.8BR, and KL.
The beam power was 24~kW as of 2013,
which corresponds to $3\times 10^{13}$ ppp.
To make primary proton beam available to E16 experiment,
the high momentum beam line which is called LINE-B
is being constructed.
LINE-B borrows a small piece of the beam ($\sim 10^{-4}$) in LINE-A 
with a Lambertson-type magnet at the switchyard.
The beam is extracted to the south side of the Hadron Experimental Facility
where E16 spectrometer is to be built.


\subsection {E16 spectrometer}
\begin{figure}
\begin{center}
\begin{minipage}{0.40\linewidth}
\includegraphics[width=1.0\linewidth]{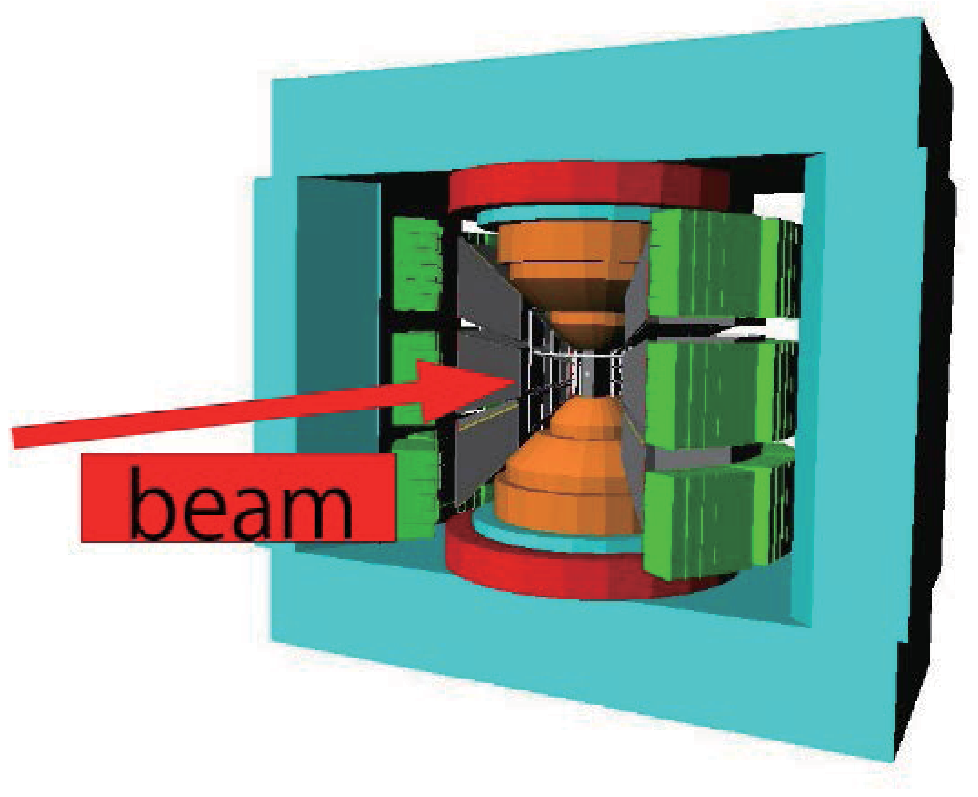}
\end{minipage}
\begin{minipage}{0.55\linewidth}
\includegraphics[width=1\linewidth,clip]{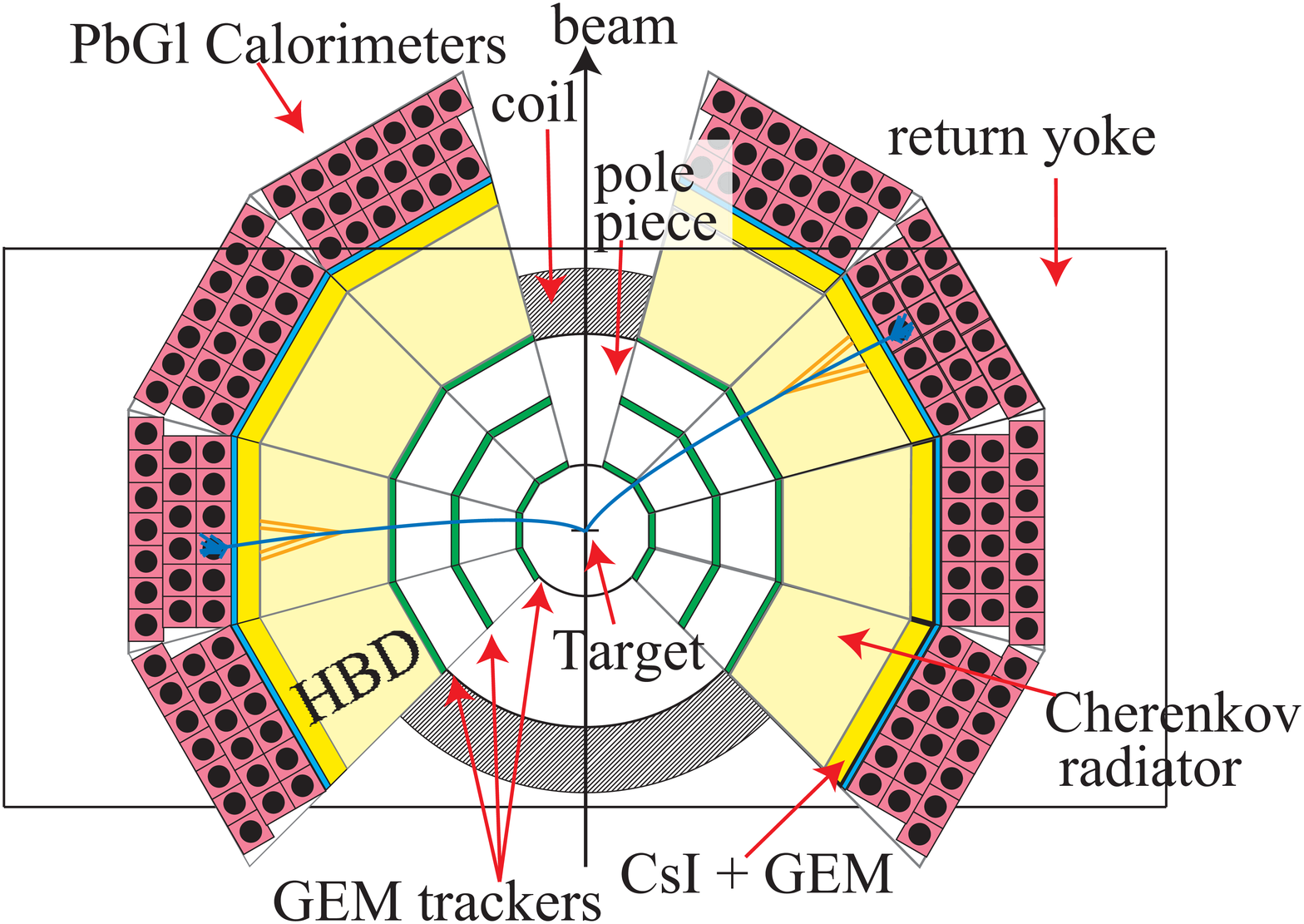}
\end{minipage}
\caption{3D view (Left) and plan view (Right) of the E16 spectrometer.}
\label{fig:spectrometer}
\end{center}
\end{figure}

A 3D view of the J-PARC E16 spectrometer is shown on the left side
of Fig.~\ref{fig:spectrometer}.
The E16 detectors are all installed inside a giant dipole magnet
with a field strength of 1.7~T at the center.
A horizontal cut view at the center is presented on the right side of Fig.~\ref{fig:spectrometer}.
The proton beam runs from bottom to the top of the figure,
and hit the target at the center of the spectrometer.
The spectrometer consists of GEM Trackers (GTR)~\cite{GTR},
Hadron Blind Detectors (HBD)~\cite{HBD}, and lead glass (LG) calorimeters.

A module is defined as a set of GTR, HBD and LG which covers 30 degrees
both horizontally and vertically. The full design of the spectrometer
consists of 26 modules. GTR is made of three layers of position-sensitive
GEM tracking chambers
with the sizes of 100 $\times$ 100 mm$^2$, 200 $\times$ 200 mm$^2$ and 300 $\times$ 300 mm$^2$,
respectively. HBD is a cherenkov counter and is used for electron identification
together with LG.
Particle tracks in the magnetic field 
are reconstructed with GTR so that the momenta are measured.
Electron candidates are selected with HBD and LG.
Position resolution of 100~$\mu m$ with incident angles of up to 30 degrees
is required for GTR.
Rejection factors of 100 and 25 are required for HBD and LG, respectively.

\begin{figure}
\begin{center}
\begin{minipage}{0.45\linewidth}
\includegraphics[width=0.9\linewidth]{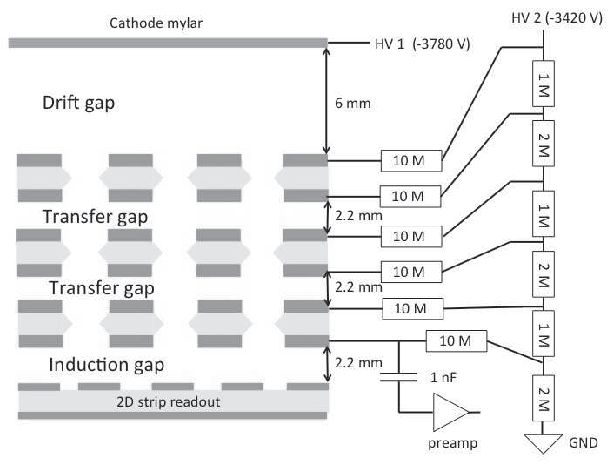}
\caption{Schematic of a GEM tracking chamber.}
\label{fig:gem}
\end{minipage}
\begin{minipage}{0.05\linewidth}
~
\end{minipage}
\begin{minipage}{0.45\linewidth}
\includegraphics[width=0.9\linewidth]{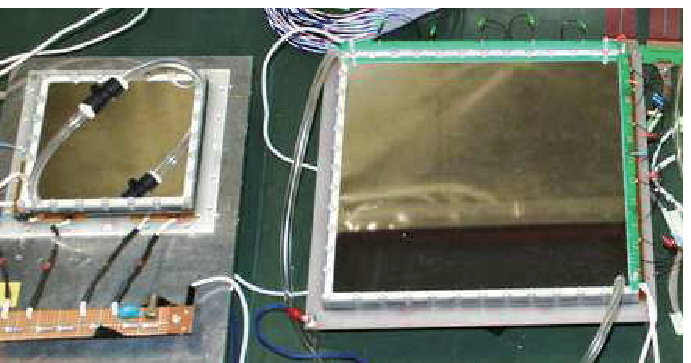}
\vspace{2ex}
\caption{Picture of the production type of the GEM tracking chambers.
The sizes are 100 $\times$ 100 mm$^2$ and 200 $\times$ 200 mm$^2$, respectively.
}
\label{fig:picgem}
\end{minipage}
\end{center}
\end{figure}

\begin{figure}
\begin{center}
\begin{minipage}{0.45\linewidth}
\includegraphics[width=.9\linewidth]{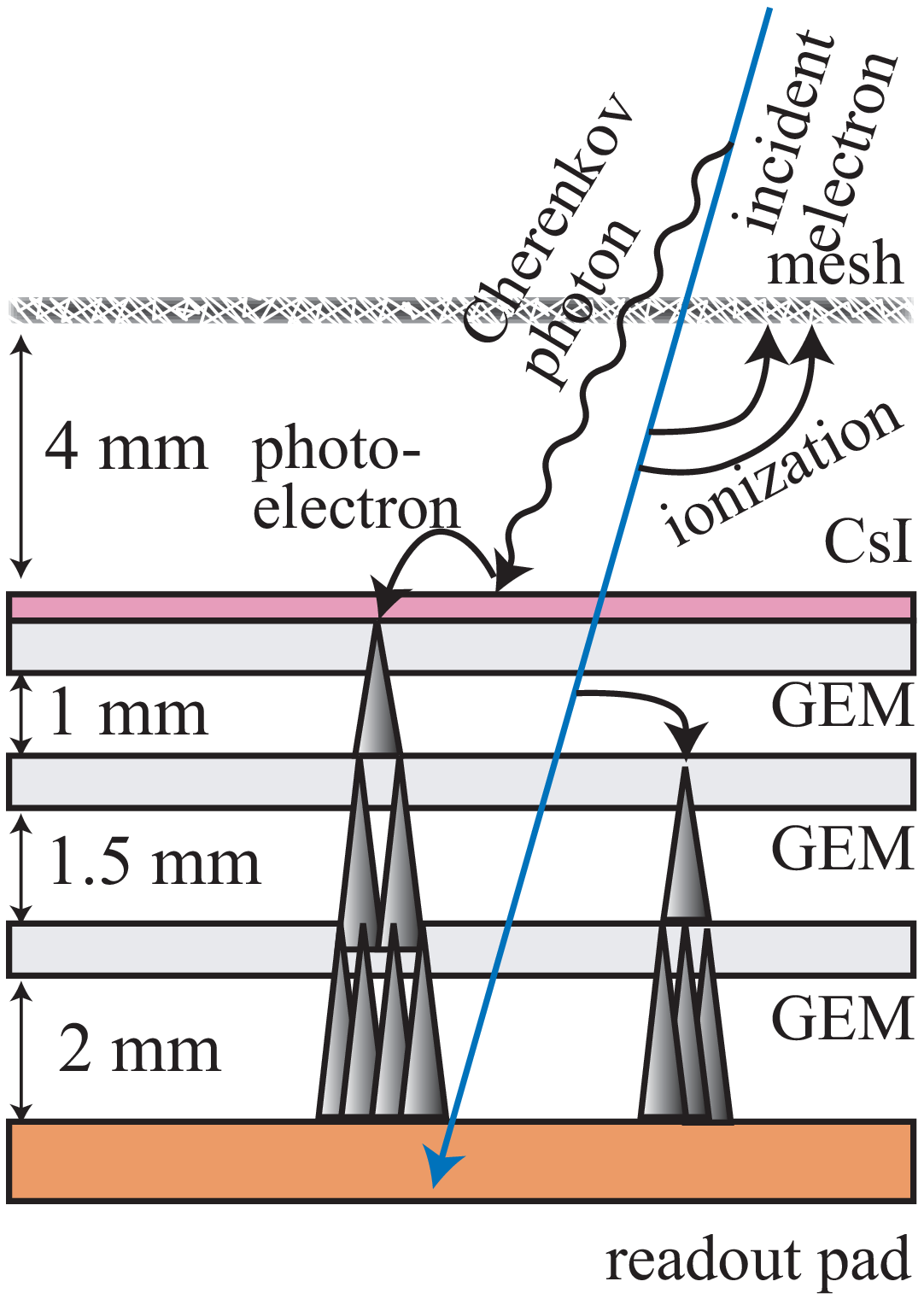} 
\caption{Schematic of the photocathode of HBD.}
\label{fig:hbd-schematics}
\end{minipage}
\begin{minipage}{0.05\linewidth}
~
\end{minipage}
\begin{minipage}{0.45\linewidth}
\vspace{10ex}
\includegraphics[width=.9\linewidth]{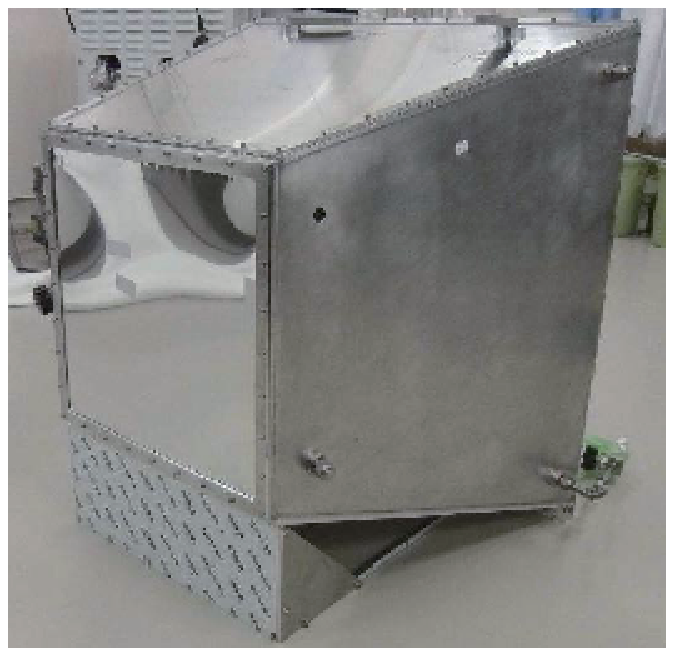}
\vspace{3ex}
\caption{Picture of a prototype of HBD in real size.}
\label{fig:hbdpic}
\end{minipage}
\end{center}
\end{figure}

\subsection{R \& D status of the spectrometer}
A schematic of a GEM tracking chamber is shown in Fig.~\ref{fig:gem}.
Ar + CO2 (70:30) is used as the amplification gas.
Three GEM foils are placed and they amplify the ionization electrons
produced by a traversing charged particle in the gap above the top GEM.
The amplified signal is readout with two dimensional strip readout board.
A custom preamp board using APV25 chip~\cite{apv} has been developed.
The mass production type of GEM tracking chambers with three different sizes
and preamp boards
have been built. The performance of them was evaluated with charged particle beams
at J-PARC and ELPH.
The required resolution of 100~$\mu m$ was
achieved for incident angles of up to 30 degrees.
A picture of the GEM chambers with two sizes are shown in Fig.~\ref{fig:picgem}.
First level trigger is readout from the bottom of the GEM foil.
A prototype of ASD (Amplifier-Shaper-Discriminator) ASIC for the trigger readout
has also been developed.

HBD is a type of cherenkov detector using
CsI evaporated GEM as a photocathode.
Our HBD has been developed based on the PHENIX HBD experience~\cite{HBDphenix}.
CF$_4$ serves as radiator and amplification gas.
With the radiator length of 50~cm, 11 photoelectrons are expected.
A schematic of the photocathode is shown in Fig.~\ref{fig:hbd-schematics}.
The incident electron emit cherenkov photons. The photons
are converted into photoelectrons by the CsI layer
which is evaporated on top of the top GEM.
The photoelectrons are then amplified by the GEMs.
A weak reverse bias field is applied in the gap between the mesh and the top GEM,
so that the ionization electrons in the gap are swept into the mesh.
Even with the reverse bias field, photoelectrons produced near the top GEM surface
are still attracted by the GEM's field and are amplified to be detected.
Therefore, HBD is blind to ionization while is sensitive to cherenkov photons.
The size of the photocathode of a HBD module is 600 $\times$ 600 mm$^2$ and
four photocathodes with a size of 300 $\times$ 300~mm$^2$ are used to fill the module.
Extensive R \& D effort has been performed to establish HBD components
such as efficient and robust CsI GEMs, airtight chambers and readout boards.
Prototypes of GEMs with and without CsI, chambers and readout boards in small sizes 
and real sizes were produced.
A picture of a prototype of the HBD chamber in real size is shown in Fig.~\ref{fig:hbdpic}.
It corresponds to a module of the spectrometer.
A beam test was performed with negatively charged particle beam of
1.0~GeV/$c$ at the J-PARC
K1.1BR to evaluate the performance of a prototype HBD in small size.
A pion rejection factor of 100 with an electron efficiency of 80\% was achieved
using cluster size analysis.
The prototype of HBD in real size also operates well.
The performance meet the required rejection and efficiency for the experiment.

\subsection{Schedule}
The experiment was approved as stage-1 in 2007.
Detector R\&D started in 2008.
The construction budget of the high-p beam line was approved in 2013.
Technical design report was submitted and the mass production
of detectors started in 2014. 
Due to the budgetary limitation, we start with one third of the full design.
The one third of the full design will be ready for the first physics
run which is anticipated in JFY2016.

\section{Other related experiments at J-PARC}
J-PARC E26 experiment has been proposed to investigate
$\omega$ meson in nuclear medium\cite{Ozawa1, Ozawa2}.
It plan to use $\pi^-$ beam at J-PARC K1.8 beam line
with a momentum of 1.8~GeV/$c$ and with an intensity of $1\times 10^7$ / pulse.
The reaction $\pi^{-} A \rightarrow \omega n X$ is used.
Invariant mass of $\omega$ meson is measured
with $\omega \rightarrow \pi^0\gamma \rightarrow 3\gamma$ decay mode.
When neutron is detected at zero degree, recoilless $\omega$ production
is realized. 
The condition is suitable for the study of in-medium effect.
Nuclear $\omega$ bound state can be searched via 
forward neutron measurement.

J-PARC E29 experiment has been proposed to investigate in-medium mass modification
of $\phi$ meson via $\phi$ meson bound state in target nucleus\cite{Ohnishi1,Ohnishi2}.
It plan to use $\bar{p}$ beam with a momentum of 1.1~GeV/$c$
and with an intensity of $1\times 10^6$/pulse.
When four strangeness are identified in the final state,
the double $\phi$ production, $\bar{p} + p \rightarrow \phi \phi$, dominates.
The forward-going $\phi$ meson is detected via $K^+K^-$ decay.
The $\phi$ meson in nucleus is detected via $\Lambda K^+$ decay,
which occur only when $\phi$ is in nucleus ($\phi+p \rightarrow \Lambda K^+$).
Missing mass spectrum is calculated with the beam momentum
and the forward-going $\phi$ momentum.
The backward $\phi$ is at the same order of Fermi momentum which is detected
via $\Lambda K^+$ decay in nucleus.

When high intensity high resolution secondary beam line (HIHR)
which was proposed by RCNP is realized, 
experimental study of $\phi$ meson in nuclear medium 
using a similar method as J-PARC E26 can be done.
A $10^9/$pulse $\pi^-$ beam with a momentum of $\sim 2$~GeV is used
to induce the $\pi^- + p \rightarrow \phi n$ reaction.
If the neutron is identified at the forward angle, ultra slow $\phi$ is selected.
Forward neutron measurement may lead to observation of
nuclear $\phi$ bound state.
About 10 times more $\phi$ compared to E16 is expected to be collected
with $\beta\gamma<0.5$.

\section{Summary}
The origin of hadron mass is studied through mass modification of vector mesons.
There are many measurements of dilepton invariant mass in hot and cold system.
There exists some modification but the origin is not yet clear.
J-PARC E16 experiment pursue it by collecting 100 times more statics
compare to the KEK E325 experiment.
We expect to obtain double peak structure in $\phi$ meson invariant mass spectra,
wide range of system size dependence of the in-medium modification,
and the dispersion relation of $\phi$ meson in nuclear medium.
We start with one-third of the design configuration and physics
run is anticipated in JFY2016.
More experiments regarding vector meson mass modification are planned and
whole together provide further insights on the origin of mass,
and the chiral symmetry.

\section{Acknowledgments}
We would like to give our thanks to the staff of KEK Fuji test beam line, ELPH at Tohoku
University, LEPS at SPring-8, J-PARC Hadron Experimental Facility and
RIKEN RI Beam Factory for their support for the beam test of detectors.
We also would like to thank to KEK electronics system group (e-sys)
for their help in the development and test of the readout circuits.
This study was partly supported by 
Grant-in-Aid for JSPS Fellows 12J01196, RIKEN SPDR program,
and MEXT/JSPS KAKENHI Grant Numbers 19654036, 19340075, 21105004 and 26247048.

\end{document}